\documentclass[conference]{IEEEtran}
\IEEEoverridecommandlockouts
\usepackage{cite}
\usepackage{amsmath,amssymb,amsfonts}
\usepackage{algorithmic}
\usepackage{graphicx}
\usepackage{textcomp}
\usepackage{xcolor}
\usepackage{CJKutf8}
\usepackage{tabularray}
\usepackage{subfigure}
\usepackage{caption}
\usepackage{subcaption}
\usepackage{verbatim}
\usepackage{subfig}
\usepackage{bm}
\usepackage{tabularx}

\def\BibTeX{{\rm B\kern-.05em{\sc i\kern-.025em b}\kern-.08em
    T\kern-.1667em\lower.7ex\hbox{E}\kern-.125emX}}

\begin{document}

\columnsep 0.241in
\topskip 0.19in

\begin{CJK}{UTF8}{gbsn}

\title{CASC: Condition-Aware Semantic Communication with Latent Diffusion Models}

\author{\IEEEauthorblockN{Weixuan Chen and Qianqian Yang\textsuperscript{\textsection}}
\IEEEauthorblockA{{
}
{College of Information Science and Electronic Engineering, Zhejiang University, Hangzhou 310007, China}\\
\{12231075, \textsuperscript{\textsection}qianqianyang20\}@zju.edu.cn}
}

\maketitle

\begin{abstract}

Diffusion-based semantic communication methods have shown significant advantages in image transmission by harnessing the generative power of diffusion models. However, they still face challenges, including generation randomness that leads to distorted reconstructions and high computational costs. To address these issues, we propose CASC, a condition-aware semantic communication framework that incorporates a latent diffusion model (LDM)-based denoiser. The LDM denoiser at the receiver utilizes the received noisy latent codes as the conditioning signal to reconstruct the latent codes, enabling the decoder to accurately recover the source image.
By operating in the latent space, the LDM reduces computational complexity compared to traditional diffusion models (DMs). Additionally, we introduce a condition-aware neural network (CAN) that dynamically adjusts the weights in the hidden layers of the LDM based on the conditioning signal. This enables finer control over the generation process, significantly improving the perceptual quality of the reconstructed images.
Experimental results show that CASC significantly outperforms DeepJSCC in both perceptual quality and visual effect. Moreover, CASC reduces inference time by 51.7\% compared to existing DM-based semantic communication systems, while maintaining comparable perceptual performance. The ablation studies also validate the effectiveness of the CAN module in improving the image reconstruction quality. 
\end{abstract}

\begin{IEEEkeywords}
Semantic communications, condition-aware weight control, latent diffusion models, image transmission
\end{IEEEkeywords}

\section{Introduction}



As the volume of wireless data in daily life continues to grow, there is an increasing demand for more efficient and reliable communication schemes. Traditional communication systems, however, often introduce redundancy to ensure bit-level accurate transmission \cite{shannon1948mathematical}. In contrast, semantic communication (SemCom) \cite{yang2022semantic} focuses on the extraction and transmission of data semantics, significantly enhancing transmission efficiency. Bourtsoulatze et al. \cite{bourtsoulatze2019deep} were the first to propose the deep joint source and channel coding (DeepJSCC) scheme for wireless image transmission, a key technology in SemCom. This scheme leverages a convolutional neural network (CNN) for joint source and channel coding, surpassing the performance of traditional separation-based methods.

Inspired by the success of DeepJSCC, numerous subsequent studies have aimed to enhance its performance. Kurka et al. \cite{kurka2020deepjscc} proposed DeepJSCC-f, which incorporates channel output feedback and transmits images progressively through multiple refinement layers. Zhang et al. \cite{zhang2022wireless} introduced MLSC-image, a method that separately extracts low-level and high-level semantic information from images, leveraging high-level semantics to improve image reconstruction performance. Zhang et al. \cite{zhang2023predictive} proposed a SemCom method capable of predicting the performance of the transmission task based on channel conditions, compression ratio, and image content, and automatically adjusting the compression ratio to enhance transmission efficiency. Despite these advancements, methods relying on a CNN-based autoencoder architecture for image reconstruction often achieve limited performance improvement over traditional methods, particularly in the high compression ratio regime.

Han et al. \cite{han2022semantic} were among the first to demonstrate the potential of generative models by developing a SemCom system for speech transmission. This system utilizes a pretrained Generative Adversarial Network (GAN) to reconstruct high-quality speech from essential transmitted features, requiring only 0.2\% of the symbols used by existing methods, while preserving comparable quality. With the advancement of generative models, several studies have also leveraged their capabilities for the efficient transmission of images. For facial images, Han et al. \cite{han2023generative} proposed a SemCom framework based on the semantic StyleGAN, achieving an exceptionally low compression ratio of 1/3072, which significantly boosts transmission efficiency.
%
Huang et al. \cite{huang2021deep} employed adversarial training between a generator and a discriminator to achieve visually appealing image reconstructions. Jiang et al. \cite{jiang2024diffsc} introduced DiffSC, a system that uses a denoiser based on the diffusion probabilistic model (DPM) to reconstruct high-quality images by conditioning on the output of the DeepJSCC decoder. This approach resulted in a 5\%-10\% improvement in the PSNR of reconstructed images compared to DeepJSCC.
Grassucci et al. \cite{grassucci2023generative} proposed a diffusion-based SemCom method that transmits only the layout information of cityscape images and applies fast denoising to the received noisy semantic layout maps to reconstruct the original images.

Although diffusion model \cite{dhariwal2021diffusion} (DM)-based SemCom systems have achieved significant advances in image transmission, two major limitations remain.
First, the inherent randomness in the generation process of DM-based SemCom systems can lead to semantic loss, causing the generated content to deviate from the original one. 
To address this, it is crucial to introduce a control mechanism to the generation process. 
One potential way is to manipulate the weights of the DM based on the transmitted semantic information to flexibly control the generation process, thereby enhancing the reconstruction performance.
Secondly, current DM-based SemCom systems are not sufficiently lightweight for efficient deployment. Unlike other generative models that operate in the latent domain, DMs typically function in the image domain, which results in high computational costs and overall latency.
%




\begin{figure}[htbp]
\begin{center}
\centerline{\includegraphics[width=1\linewidth]{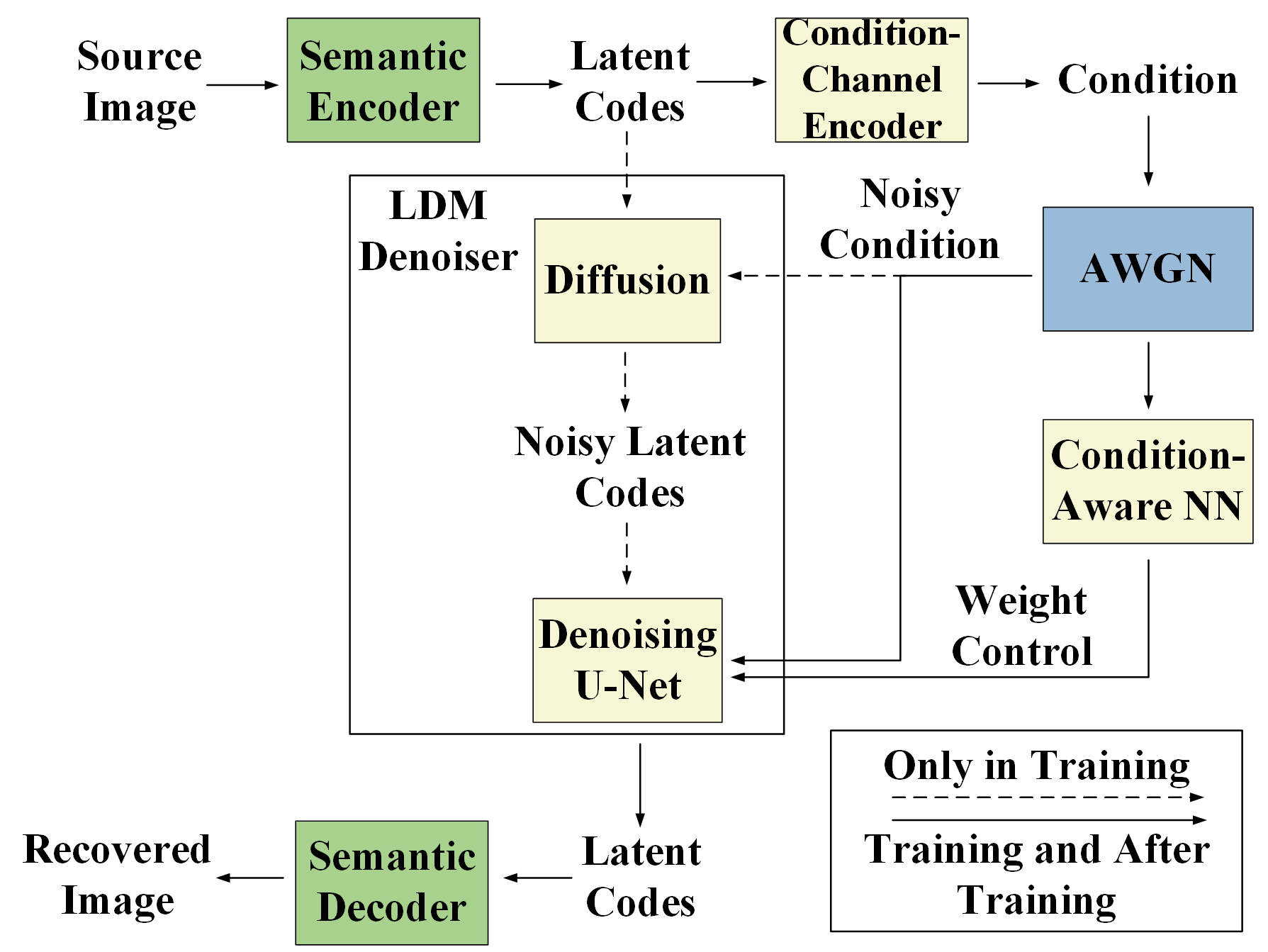}}
\caption{Our proposed condition-aware SemCom framework with latent diffusion models.}
\label{fig.1}
\end{center}
\vskip -0.1in
\end{figure}

In this paper, we propose CASC, a condition-aware semantic communication framework that integrates a latent diffusion model (LDM) \cite{rombach2022high} denoiser and a condition-aware neural network (CAN) \cite{cai2024condition}. The LDM denoiser improves image reconstruction at the receiver by using compressed noisy latent codes as the conditioning signal, guiding the generation process to produce more accurate outputs. By operating in the latent domain rather than the image domain, this denoiser also reduces computational demands.
%
%
Moreover, to further enhance the fidelity of the reconstructed images, CASC introduces CAN, which assigns dynamic weights to specific layers of the LDM denoiser based on the received condition signal, effectively controlling the generation process. Experimental results show that CASC significantly outperforms DeepJSCC in perceptual quality and visual effect. Additionally, CASC reduces inference time by 51.7\% compared to the existing diffusion-based SemCom approach, DiffSC, while maintaining comparable performance.


\section{System Model}



In this section, we introduce the system model of our proposed CASC, as shown in Fig.~\ref{fig.1}. 
CASC consists of a semantic encoder and a condition-channel encoder at the transmitter, as well as a condition-aware neural network (CAN), a denoising U-Net (denoiser), and a semantic decoder at the receiver.
At the transmitter, the semantic encoder extracts latent codes (i.e., semantic information) from the source image. 
These latent codes are then compressed into a condition signal by the condition-channel encoder and transmitted to the receiver over an additive white Gaussian noise (AWGN) channel.
At the receiver, the condition-aware neural network generates dynamic weights based on the received noisy condition signal to control the generation process of the denoiser. 
The denoiser then reconstructs the latent codes using the noisy condition signal and dynamic weights. 
Finally, the semantic decoder uses the recovered latent codes to reconstruct the source image. 
The detailed process of image transmission is described below.

At the transmitter, the source image $\textbf{X} \in \mathbb{R}^{H \times W \times 3}$, where $H$ and $W$ represent the image height and width, respectively, is fed into the semantic encoder $\mathcal{E}( \cdot )$, which extracts the semantic information, resulting in the latent codes $\textbf{Z}$, denoted by 
\begin{equation}
    \textbf{Z} = \mathcal{E}(\textbf{X}).
\end{equation}
Subsequently, the condition-channel encoder $f( \cdot )$ encodes the latent codes $\textbf{Z} $ into the condition signal $\textbf{C} \in \mathbb{R}^{d}$, denoted by
\begin{equation}
    \textbf{C} = f(\textbf{Z}).
\end{equation}
The condition signal $\textbf{C}$ is then transmitted to the receiver over the AWGN channel. 
The received noisy condition signal $\hat{\textbf{C}}$ is mathematically represented as
\begin{equation}
    \hat{\textbf{C}} = \textbf{C} + \textbf{n},
\end{equation}
where $\textbf{n} \sim \mathcal{CN} (0, {\sigma^{2}} )$ represents the AWGN with a variance of $\sigma^{2}$.

At the receiver, the noisy condition signal $\hat{\textbf{C}}$ is fed into the condition-aware neural network $f_{\rm CAN}( \cdot )$, which generates dynamic weights $\textbf{W}$, denoted by
\begin{equation}
    \textbf{W} = f_{\rm CAN}(\hat{\textbf{C}}).
\end{equation}
These dynamic weights are used to control the generation process of the denoiser, enhancing the denoising performance.

In the diffusion process (only in training), Gaussian noise is progressively added to the latent codes $\textbf{Z}$ to obtain the noisy latent codes $\textbf{Z}_{T}$. 
In the inference process, $\textbf{Z}_{T}$ is a randomly generated Gaussian noise. 
The input to the denoising U-Net $U( \cdot )$  consists of the noisy condition signal $\hat{\textbf{C}}$, the dynamic weights $\textbf{W}$, and the noisy latent codes $\textbf{Z}_{T}$. 
The denoising U-Net adjusts the weights of its hidden layers based on $\textbf{W}$ and progressively denoises the noisy latent codes $\textbf{Z}_{T}$ using $\hat{\textbf{C}}$, resulting in the denoised latent codes $\widetilde{\textbf{Z}}$, as expressed by
\begin{equation}
    \widetilde{\textbf{Z}} = U\left( \hat{\textbf{C}},\textbf{W},\textbf{Z}_{T} \right).
\end{equation}

Finally, the denoised latent codes $\widetilde{\textbf{Z}}$ are fed into the semantic decoder $\mathcal{D}(\cdot)$ to generate the recovered image $\hat{\textbf{X}} \in \mathbb{R}^{H \times W \times 3}$, denoted by
\begin{equation}
    \hat{\textbf{X}} = \mathcal{D}(\widetilde{\textbf{Z}}).
\end{equation}
The primary objective of the system is to maximize consistency between the recovered image and the source image.

\section{Proposed Method}



In this section, we provide a detailed overview of the semantic autoencoder (comprising the semantic encoder and decoder), condition-channel encoder, denoiser, conditional-aware neural network, and the training strategy.

\subsection{Semantic Autoencoder}

\begin{figure}[htbp]
\begin{center}
\centerline{\includegraphics[width=1\linewidth]{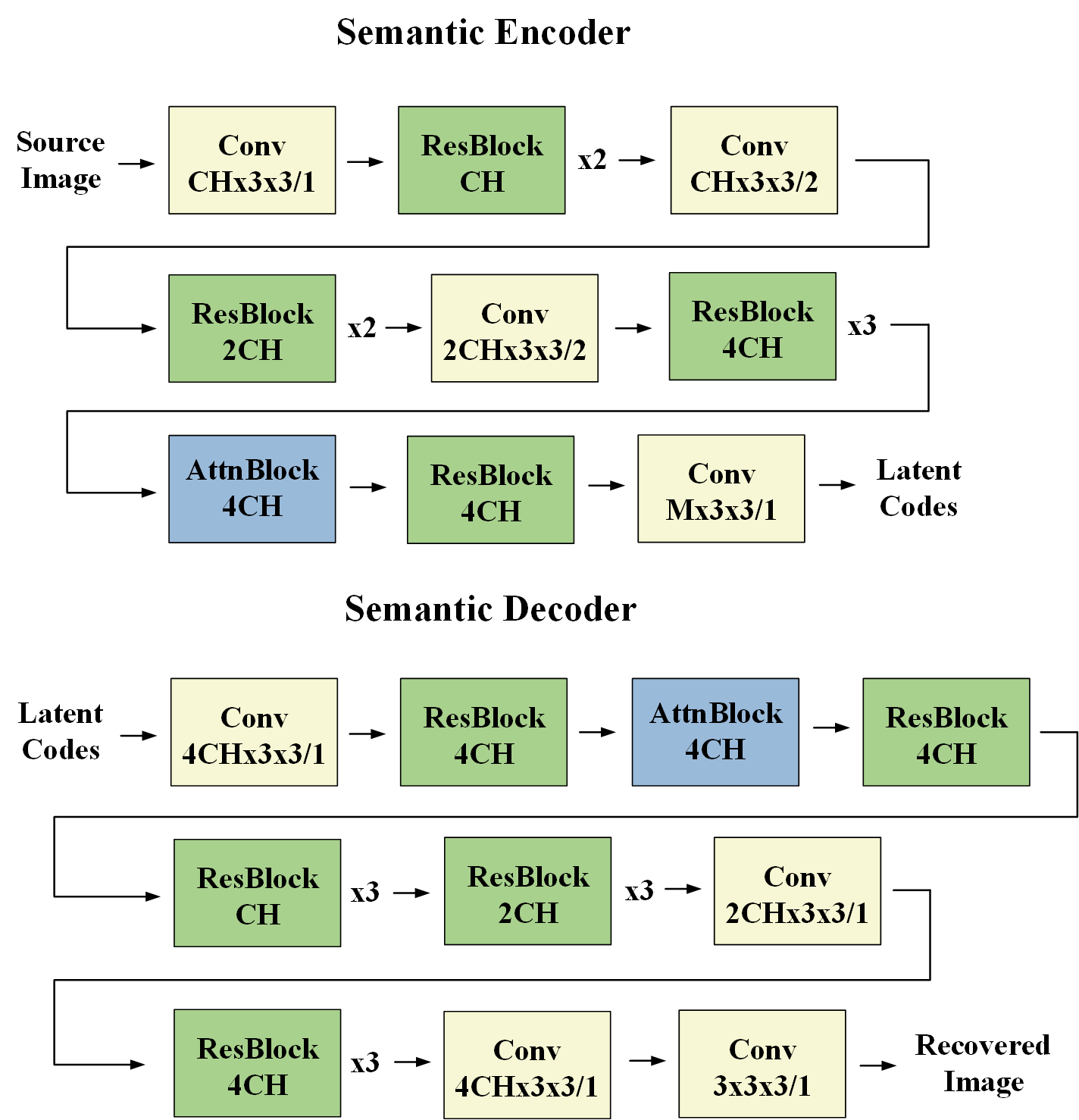}}
\caption{The network architecture of the semantic encoder and decoder.}
\label{fig.semantic_autoencoder}
\end{center}
\vskip -0.1in
\end{figure}

The semantic autoencoder consists of two components, the semantic encoder and the semantic decoder. 
The semantic encoder encodes the source image into latent codes, extracting crucial semantic information. 
The semantic decoder reconstructs the source image from the latent codes.
The network architecture of the semantic autoencoder is shown in Fig.~\ref{fig.semantic_autoencoder}. 
The numbers below the convolutional layers indicate their configurations. For example, in $CH \times 3 \times 3/1$, $CH$ represents the number of output channels, $3 \times 3$ represents the size of the convolutional kernel, and $/1$ represents the stride of the convolutional layer.
The Resblock and Attnblock refer to residual blocks and self-attention blocks, respectively, with the numbers below them representing the number of output channels.
%
We employ the same loss function used in the autoencoder from \cite{rombach2022high}, which includes the VQ regularization proposed in \cite{esser2021taming}. 
The loss function for the semantic autoencoder is denoted as $\mathcal{L}_{\rm autoencoder}$.



\subsection{Condition-Channel Encoder}

The condition-channel encoder performs channel coding on the latent codes $\textbf{Z}$ to protect them against channel noise and further compresses them into the condition signal $\textbf{C} \in \mathbb{R}^{d}$.
The condition signal $\textbf{C}$ is then transmitted to the receiver through the AWGN channel.
The structure of the condition-channel encoder is straightforward, consisting of a single convolutional layer. 
This convolutional layer is configured as $L \times 3 \times 3/1$, where $L$ controls the compression ratio.

\subsection{Denoiser}
The denoiser is based on the LDM and includes a denoising U-Net \cite{ronneberger2015u} that progressively denoises the latent codes.
The denoiser operates in two stages, the forward diffusion process and the reverse inference process. 
In the forward diffusion process, the initial latent codes $\textbf{Z}_0=\textbf{Z}$ are progressively added with Gaussian noise,
resulting in the noisy latent codes $\textbf{Z}_T$, which are then fed into the denoising U-Net. 
Here, the subscript $T$ refers to the time index.
This process occurs only in the training stage.
In the reverse inference process, a randomly generated Gaussian noise $\textbf{Z}_T$ is fed into the denoising U-Net. 
Additionally, the denoising U-Net receives the noisy condition signal $\hat{\textbf{C}}$ and the dynamic weights $\textbf{W}$ as inputs. 
The dynamic weights $\textbf{W}$ are used solely to adjust the weights of the denoising U-Net and are not directly involved in the forward diffusion or reverse inference processes.
Next, we will discuss these two processes in detail.

\subsubsection{Forward Diffusion Process}

In the forward diffusion process, we treat the noisy condition signal $\hat{\textbf{C}}$ as the known condition and the initial latent codes $\textbf{Z}$ as the target $\textbf{Z}_0$. 
The purpose of the forward diffusion process is to generate the noisy latent codes $\textbf{Z}_T$. 
Specifically, the input $\textbf{Z}_0$ is progressively added with Gaussian noise with variance $\beta_{t} \in (0,1)$ at each time index $t$,
resulting in a series of noisy latent codes from $\textbf{Z}_1$ to $\textbf{Z}_T$. This process, denoted as $q$, is mathematically expressed as 
\begin{equation}
    q\left( {\textbf{Z}_1,\dots,\textbf{Z}_T} \mid {\textbf{Z}_0} \right) = {\prod\limits_{t = 1}^{T}{q\left( \textbf{Z}_t \mid \textbf{Z}_{t-1} \right)}},
\end{equation}
\begin{equation}
    q(\textbf{Z}_t \mid \textbf{Z}_{t-1}) = \mathcal{N}\left(\textbf{Z}_t; \sqrt{1 - \beta_t} (\textbf{Z}_{t-1}), \beta_t \hat{\textbf{C}}\right).
\end{equation}


\subsubsection{Reverse Inference Process}
The reverse inference process performs the denoising function. 
During training, the denoiser learns to denoise the noisy latent codes $\textbf{Z}_T$ to compensate for the loss of semantic information caused by compression. 
This is accomplished by inferring and diffusing between the distribution of $\textbf{Z}_0$ and the distribution of $\textbf{Z}_T$. 
In this process, the noisy condition signal $\hat{\textbf{C}}$ and the noisy latent codes $\textbf{Z}_T$ are input into the denoising U-Net,
which gradually denoises $\textbf{Z}_T$ to ultimately obtain the denoised latent codes $\widetilde{\textbf{Z}}$. 
The goal is for $\widetilde{\textbf{Z}}$ to closely approximate the initial latent codes $\textbf{Z}_0$. 
The reverse process can be mathematically represented as
\begin{equation}
    p_\theta(\textbf{Z}_0, \dots, \textbf{Z}_{T-1} \mid \textbf{Z}_T) = \prod_{t=1}^{T} p_\theta(\textbf{Z}_{t-1} \mid \textbf{Z}_t),
\end{equation}
where $\theta$ represents the learnable parameters of the reverse process, and $\textbf{Z}_T$ is a randomly generated Gaussian noise, modeled as 
$p_\theta(\textbf{Z}_T) = \mathcal{N}(\textbf{Z}_T; 0, \hat{\textbf{C}})$. 
The denoising U-Net is employed to learn $p_\theta(\textbf{Z}_{t-1} \mid \textbf{Z}_t)$. 
Assuming the denoising U-Net function is $U_\theta(\textbf{Z}_t,\hat{\textbf{C}},\textbf{W},t)$, 
the loss function of the denoiser can be expressed as
\begin{equation}
    \mathcal{L}_{\rm denoiser} = \mathbb{E}_{\textbf{Z}_0, \epsilon \sim \mathcal{N}(0, 1), t} \left[ \left\| \epsilon - U_\theta(\textbf{Z}_t,\hat{\textbf{C}},\textbf{W},t) \right\|_2^2 \right],
\end{equation}
where $\epsilon$ represents a Gaussian noise variable.

\subsection{Condition-Aware Neural Network}

\begin{figure}[htbp]
\begin{center}
\centerline{\includegraphics[width=1\linewidth]{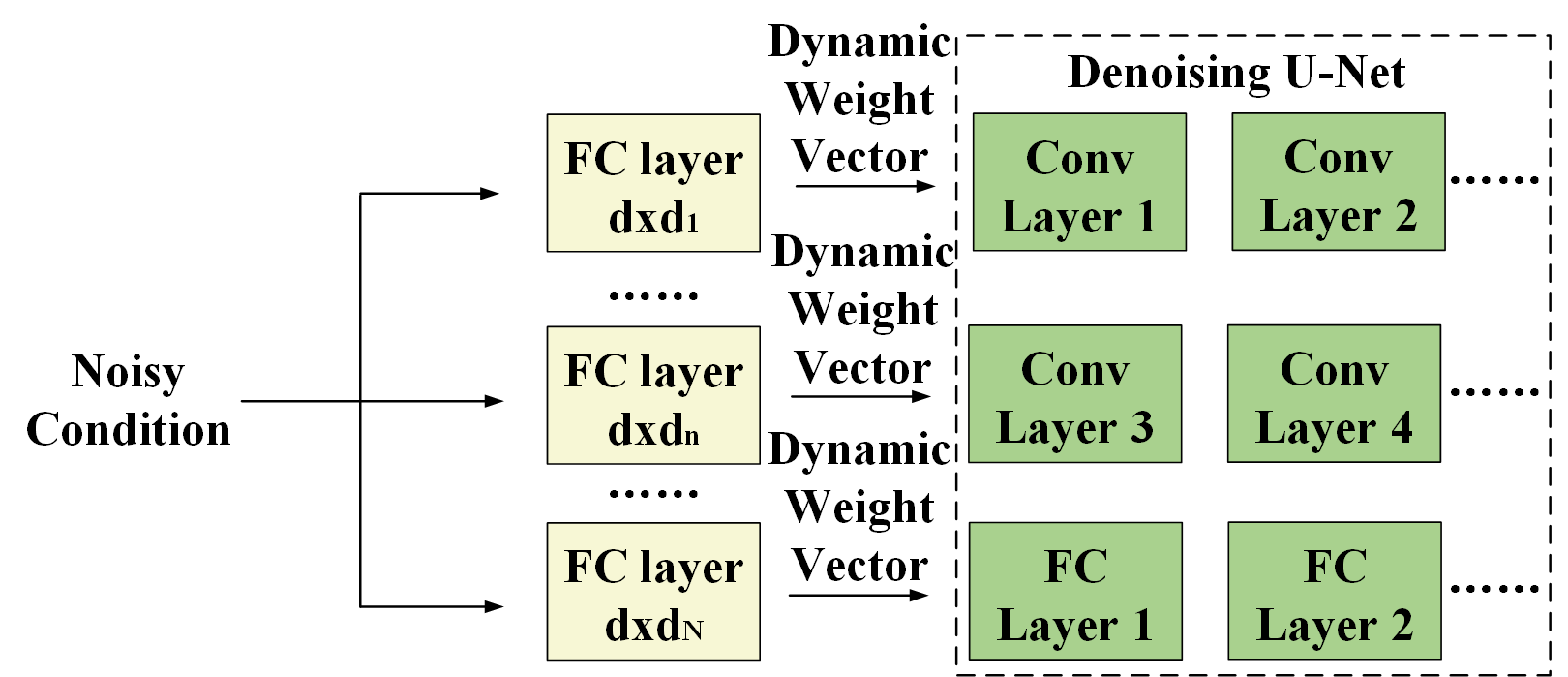}}
\caption{The process of the condition-aware neural network generating dynamic weights.}
\label{fig.CAN}
\end{center}
\vskip -0.1in
\end{figure}

The condition-aware neural network (CAN) takes the noisy condition signal $\hat{\textbf{C}}$ as input and generates dynamic weights $\textbf{W}$. 
The CAN consists of multiple parallel fully connected (FC) layers, each of which produces unique dynamic weights for different hidden layers of the denoising U-Net, as shown in Fig.~\ref{fig.CAN}.
The numbers below the FC layers represent their configurations. 
For $d \times d_n$, $d$ represents the number of input neurons, corresponding to the length of $\hat{\textbf{C}}$, 
while $d_n$ represents the number of output neurons, corresponding to the length of the dynamic weight vector.
Each dynamic weight vector is unique to each sample. 
For a single sample, the CAN can generate $N$ dynamic weight vectors, which are then applied to different hidden layers of the denoising U-Net. 
Specifically, dynamic weights are applied to selected FC and convolutional layers in the denoising U-Net. 
The length of each dynamic weight vector matches the number of weights in a corresponding hidden layer of the denoising U-Net.
In some cases, dynamic weight vectors are shared among hidden layers with identical configurations. For example, if convolutional layer 1 and convolutional layer 2 have the same number of input/output channels and the same kernel size, they can share a single dynamic weight vector.



\begin{figure}[htbp]
\begin{center}
\centerline{\includegraphics[width=0.78\linewidth]{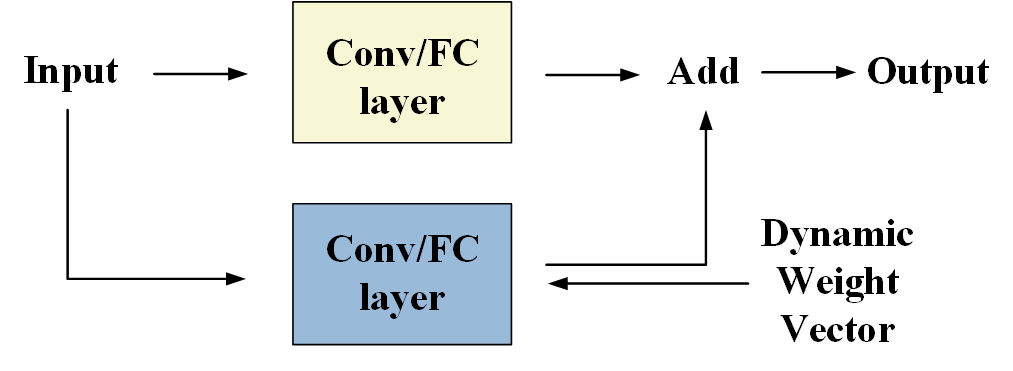}}
\caption{Illustration of the use of dynamic weights.}
\label{fig.dynamic}
\end{center}
\vskip -0.1in
\end{figure}

Next, we illustrate how the dynamic weight vectors are utilized by the hidden layers, as shown in Fig.~\ref{fig.dynamic}. 
The yellow block represents a hidden layer of the denoising U-Net, referred to as the static layer.
The blue block represents an additional hidden layer that we constructed, known as the dynamic layer.
The dynamic weight vector is applied as the weights of the dynamic layer. The input signal is then fed into both the static and dynamic layers, and their outputs are combined to produce the final output.

\subsection{Training Strategy}

We employ a two-stage training strategy to train our proposed CASC.
In the first training stage, we train only the semantic autoencoder, which includes the semantic encoder and the semantic decoder. 
The loss function in this stage is $\mathcal{L}_{\rm autoencoder}$, with an initial learning rate of $4.5 \times 10^{-6}$. 
In the second training stage, we freeze the parameters of the semantic autoencoder and train the remaining components of CASC.
The loss function in this stage is $\mathcal{L}_{\rm denoiser}$, with an initial learning rate of $1 \times 10^{-6}$.
Both stages are trained for 500 epochs.

\section{Experiments}

\begin{figure*}[htbp]
\centering
\subfigure[Results on PSNR ($\uparrow$)]{
\begin{minipage}[t]{0.325\linewidth}
\centering
\includegraphics[width=2.35in]{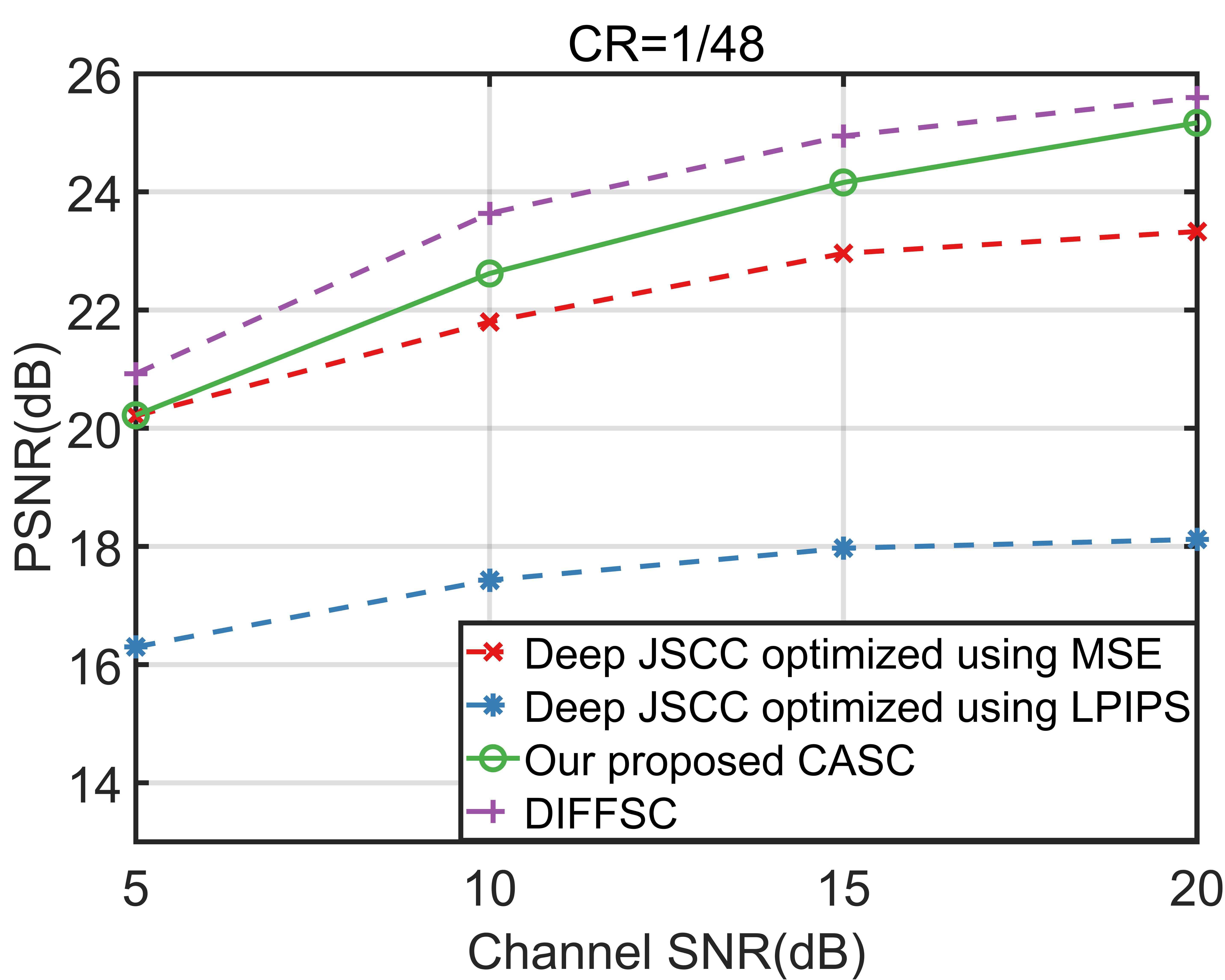}
\end{minipage}%
}%
\subfigure[Results on LPIPS ($\downarrow$)]{
\begin{minipage}[t]{0.325\linewidth}
\centering
\includegraphics[width=2.35in]{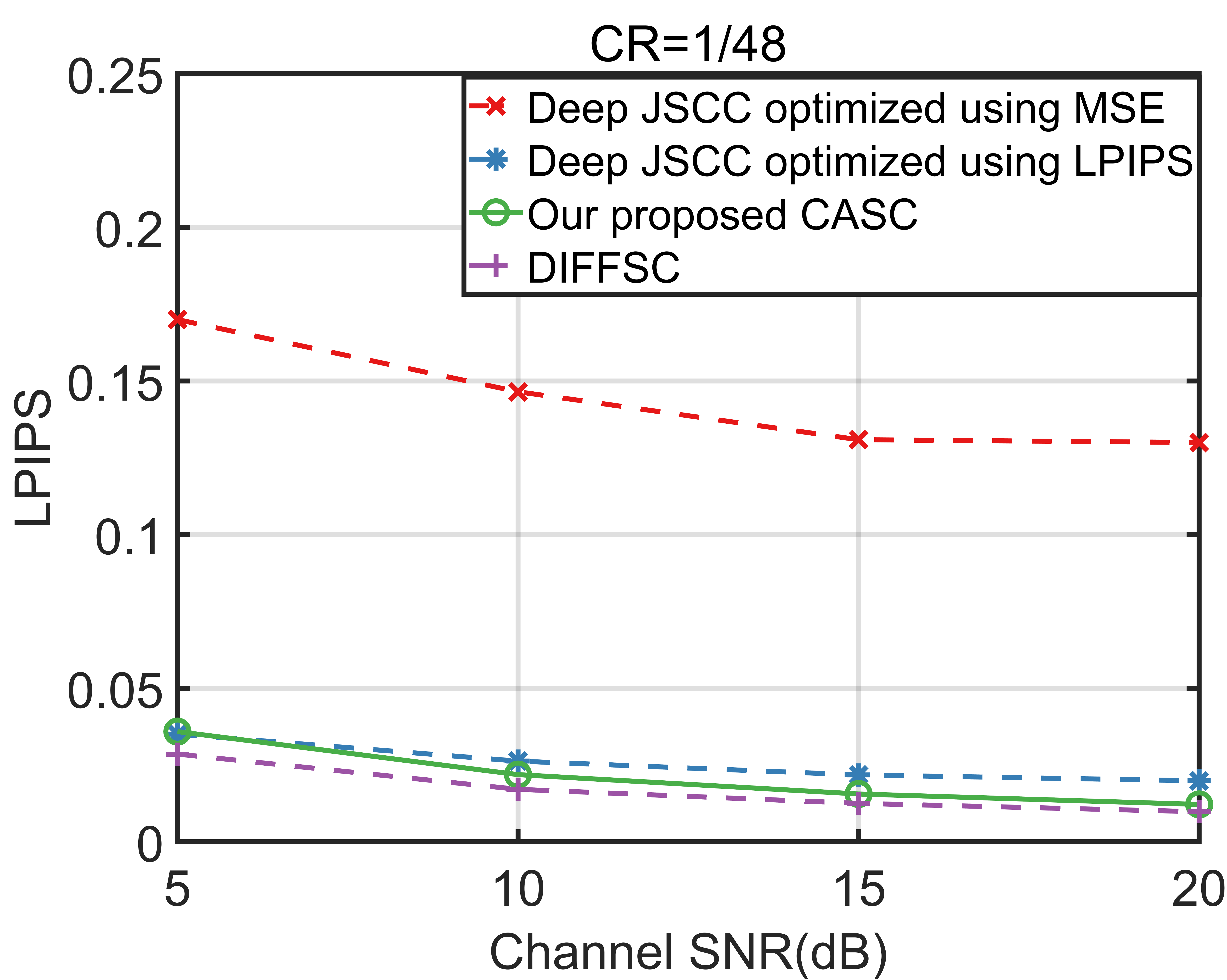}
\end{minipage}%
}%
\subfigure[Results on FID ($\downarrow$)]{
\begin{minipage}[t]{0.325\linewidth}
\centering
\includegraphics[width=2.35in]{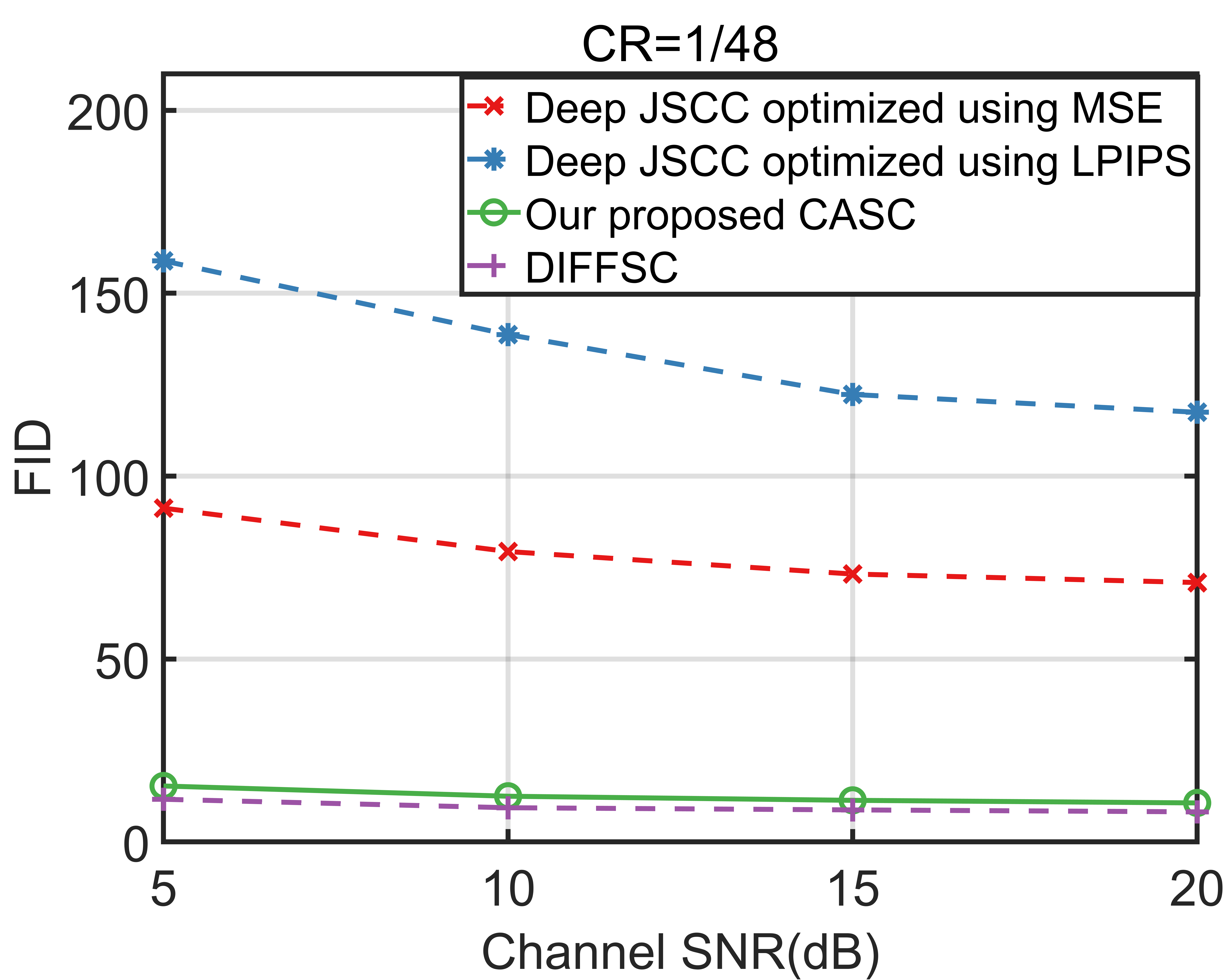}
\end{minipage}%
}%
\centering
\caption{Performance comparison at different channel SNRs with a CR of 1/48.}
\label{fig.CR=1/48}
\vskip -0.1in
\end{figure*}

\begin{figure*}[htbp]
\centering
\subfigure[Results on PSNR ($\uparrow$)]{
\begin{minipage}[t]{0.325\linewidth}
\centering
\includegraphics[width=2.35in]{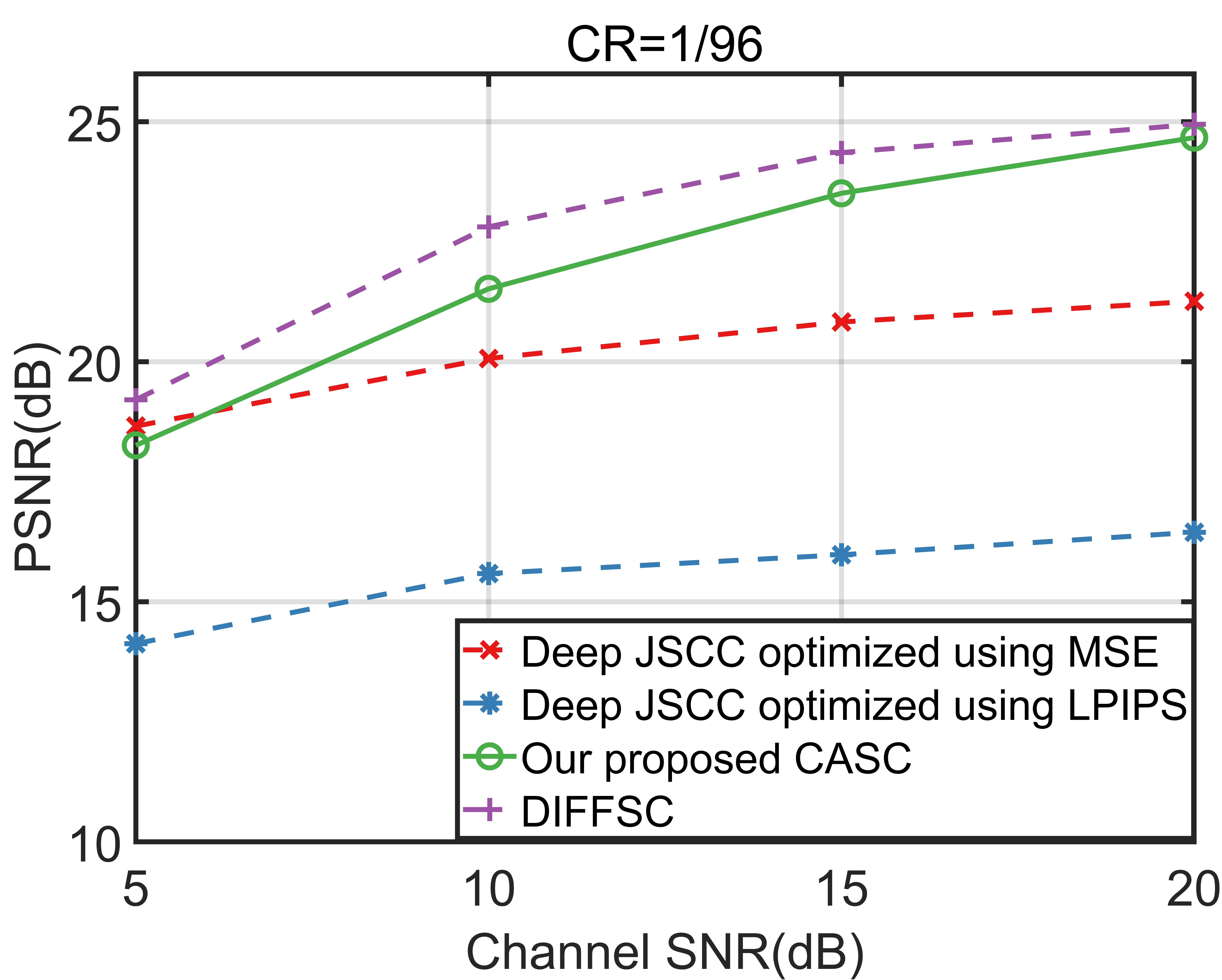}
\end{minipage}%
}%
\subfigure[Results on LPIPS ($\downarrow$)]{
\begin{minipage}[t]{0.325\linewidth}
\centering
\includegraphics[width=2.35in]{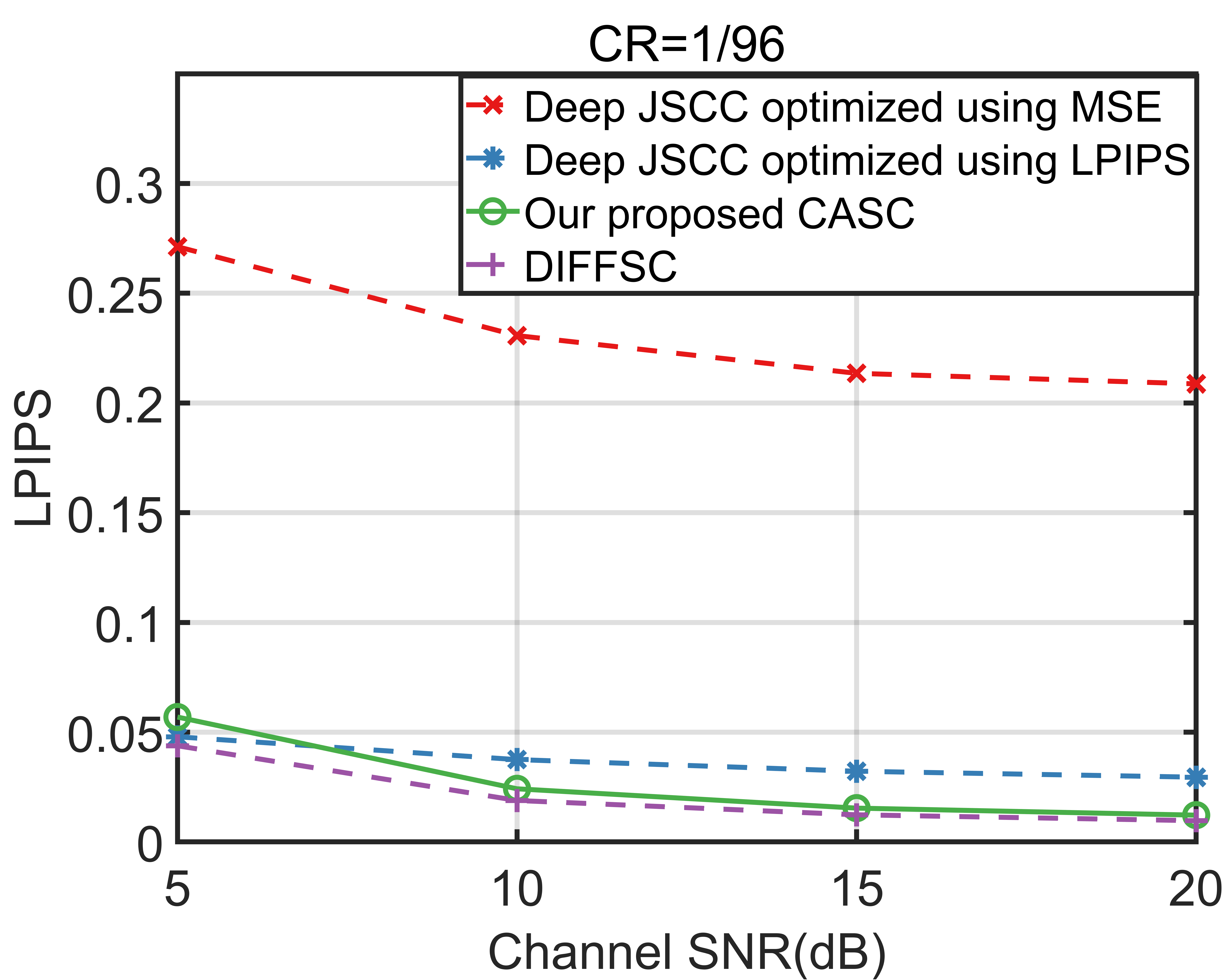}
\end{minipage}%
}%
\subfigure[Results on FID ($\downarrow$)]{
\begin{minipage}[t]{0.325\linewidth}
\centering
\includegraphics[width=2.35in]{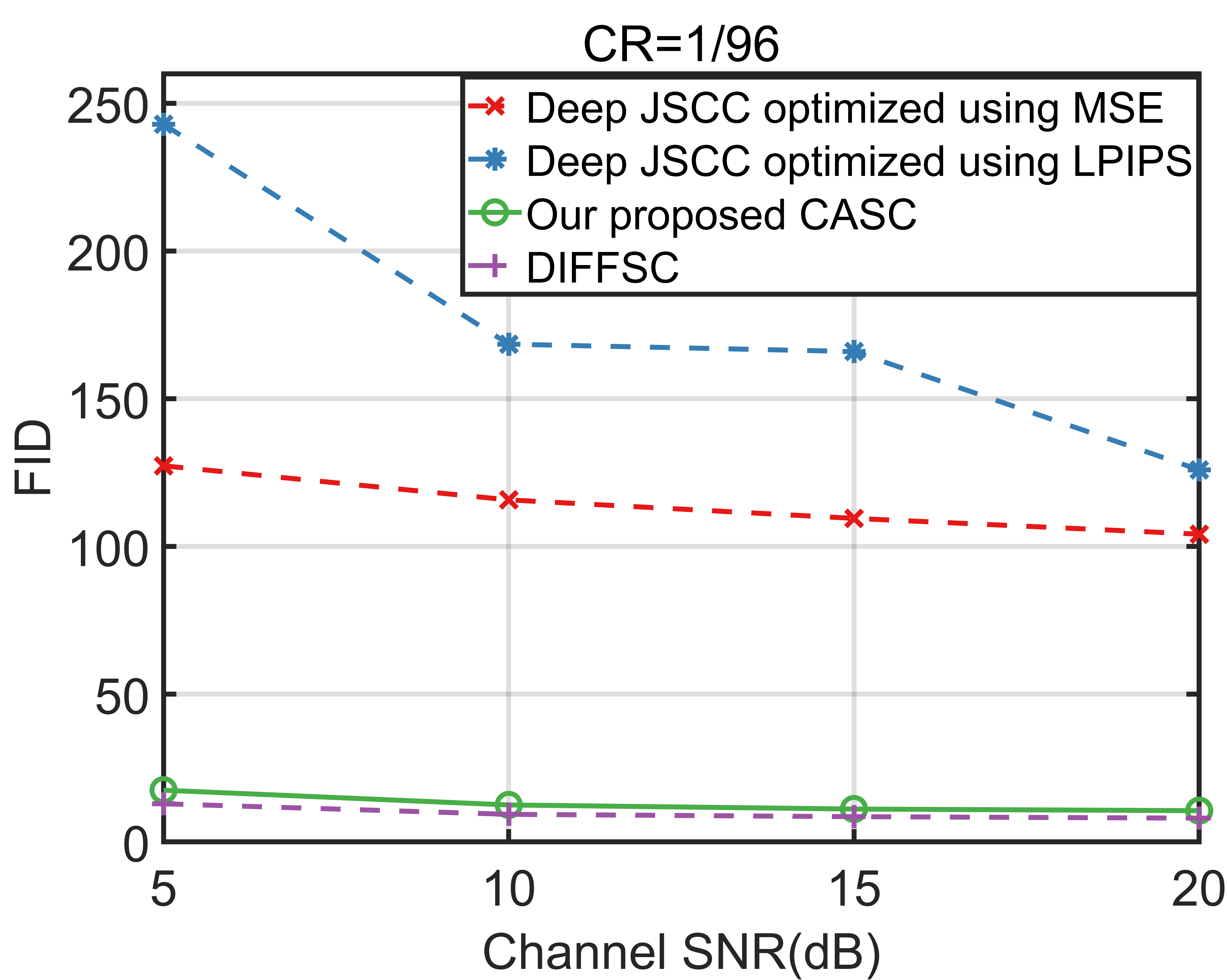}
\end{minipage}%
}%
\centering
\caption{Performance comparison at different channel SNRs with a CR of 1/96.}
\label{fig.CR=1/96}
\vskip -0.1in
\end{figure*}

\begin{figure}[htbp]
\centering
\subfigure[Source image]{
\begin{minipage}[t]{0.5\linewidth}
\centering
\includegraphics[width=0.8in]{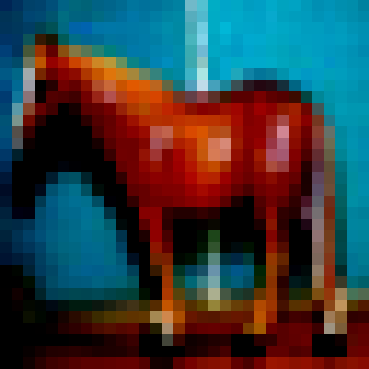}
\end{minipage}%
}%
\subfigure[Deep JSCC optimized using MSE]{
\begin{minipage}[t]{0.5\linewidth}
\centering
\includegraphics[width=0.8in]{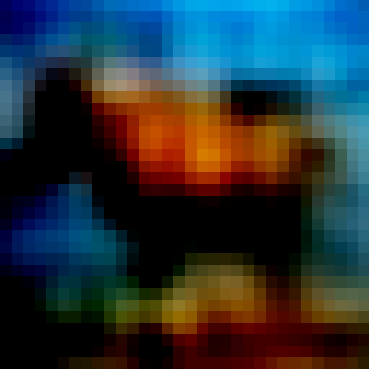}
\end{minipage}%
}%
\\
\subfigure[Deep JSCC optimized using LPIPS]{
\begin{minipage}[t]{0.5\linewidth}
\centering
\includegraphics[width=0.8in]{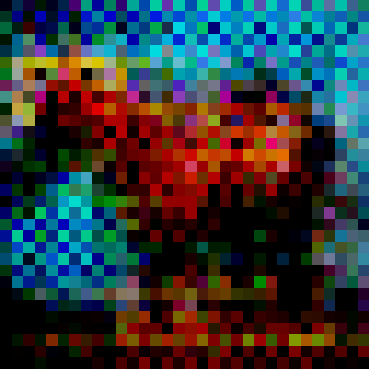}
\end{minipage}%
}%
\subfigure[Our proposed CASC]{
\begin{minipage}[t]{0.5\linewidth}
\centering
\includegraphics[width=0.8in]{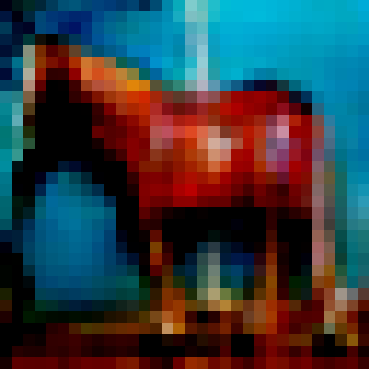}
\end{minipage}%
}%
\centering
\caption{Visual comparison at a channel SNR of 20dB with a CR of 1/48.}
\label{fig.visual}
\vskip -0.1in
\end{figure}

\begin{figure*}[htbp]
\centering
\subfigure[Results on PSNR ($\uparrow$)]{
\begin{minipage}[t]{0.325\linewidth}
\centering
\includegraphics[width=2.35in]{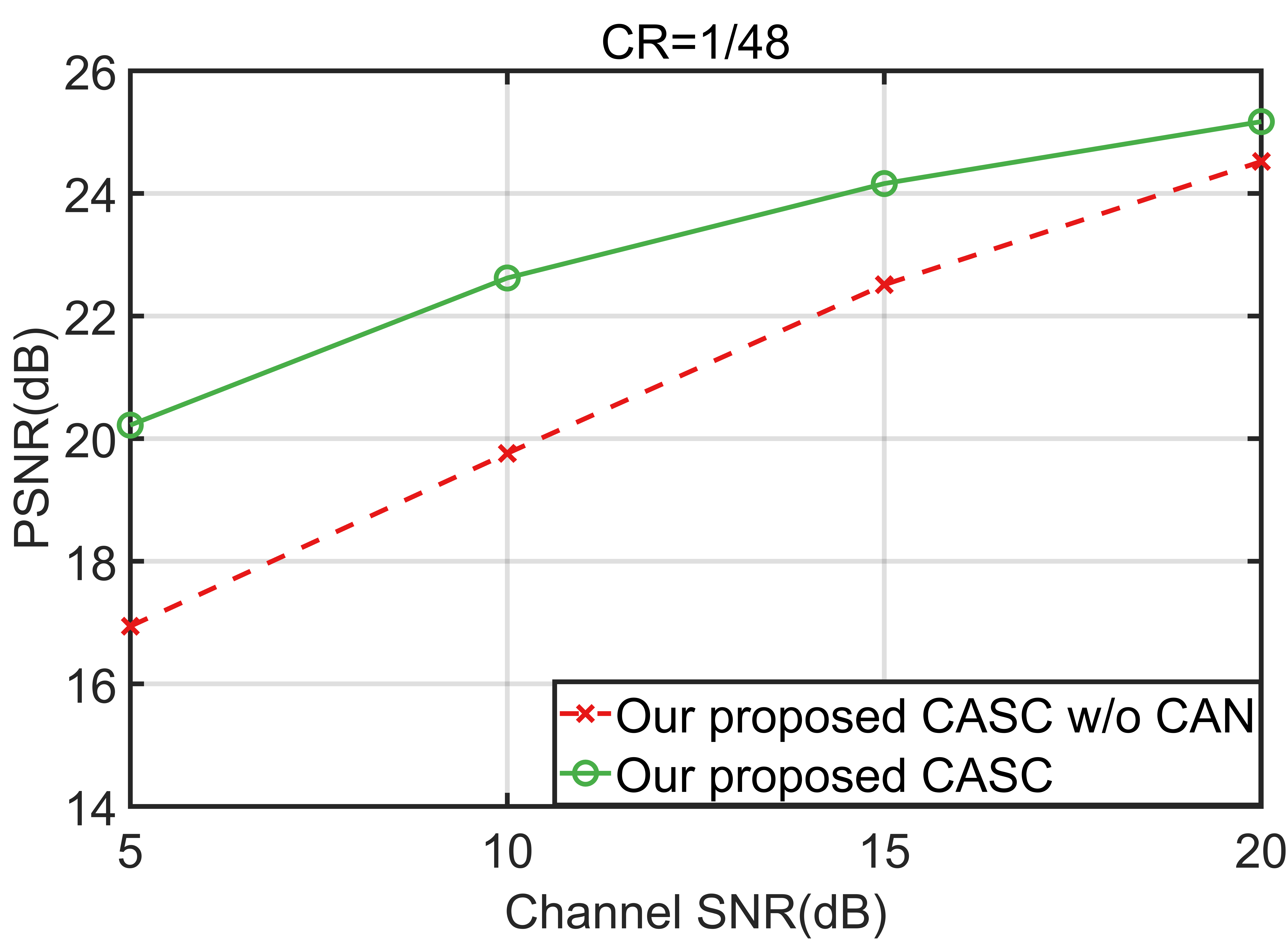}
\end{minipage}%
}%
\subfigure[Results on LPIPS ($\downarrow$)]{
\begin{minipage}[t]{0.325\linewidth}
\centering
\includegraphics[width=2.35in]{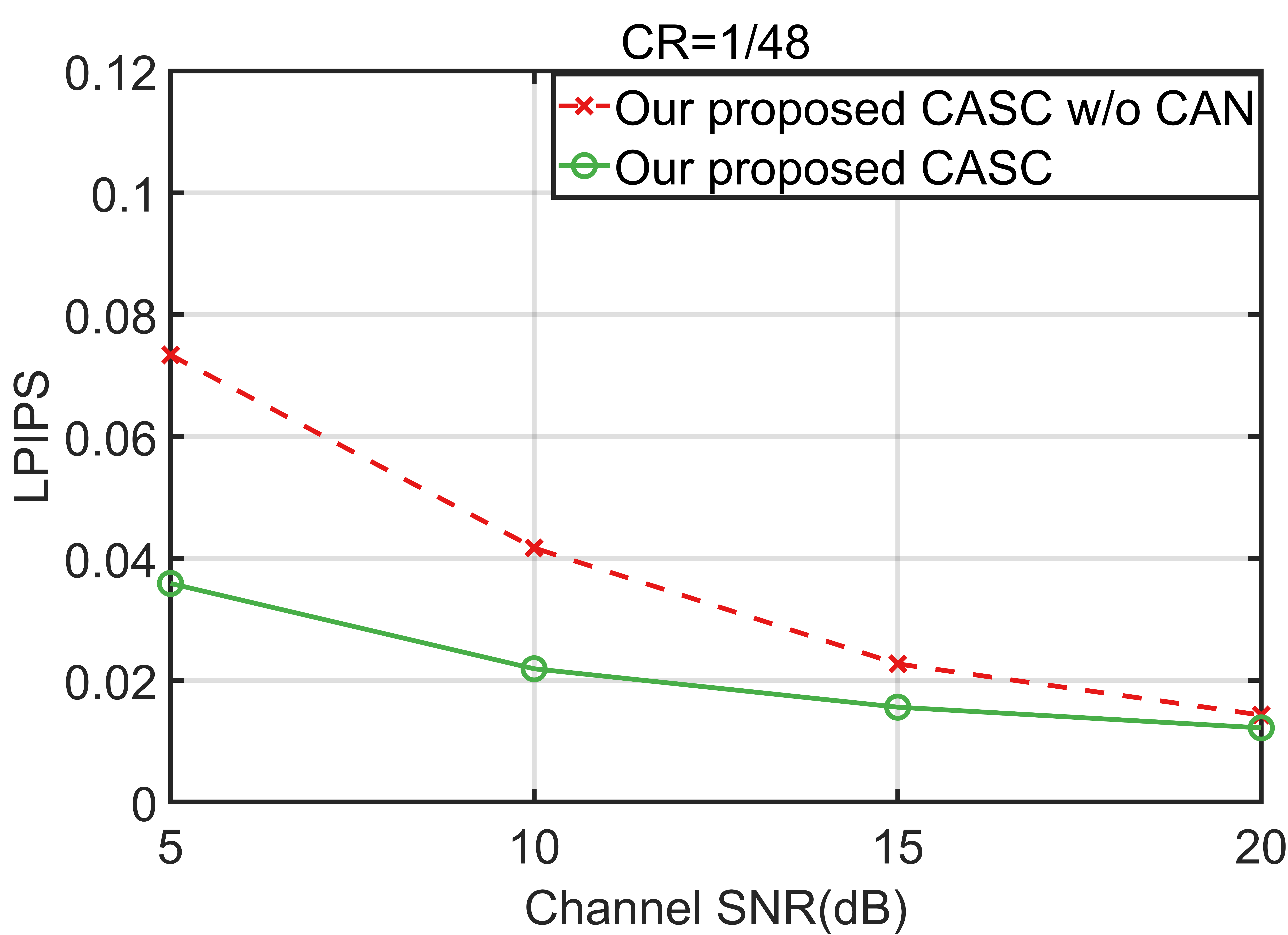}
\end{minipage}%
}%
\subfigure[Results on FID ($\downarrow$)]{
\begin{minipage}[t]{0.325\linewidth}
\centering
\includegraphics[width=2.35in]{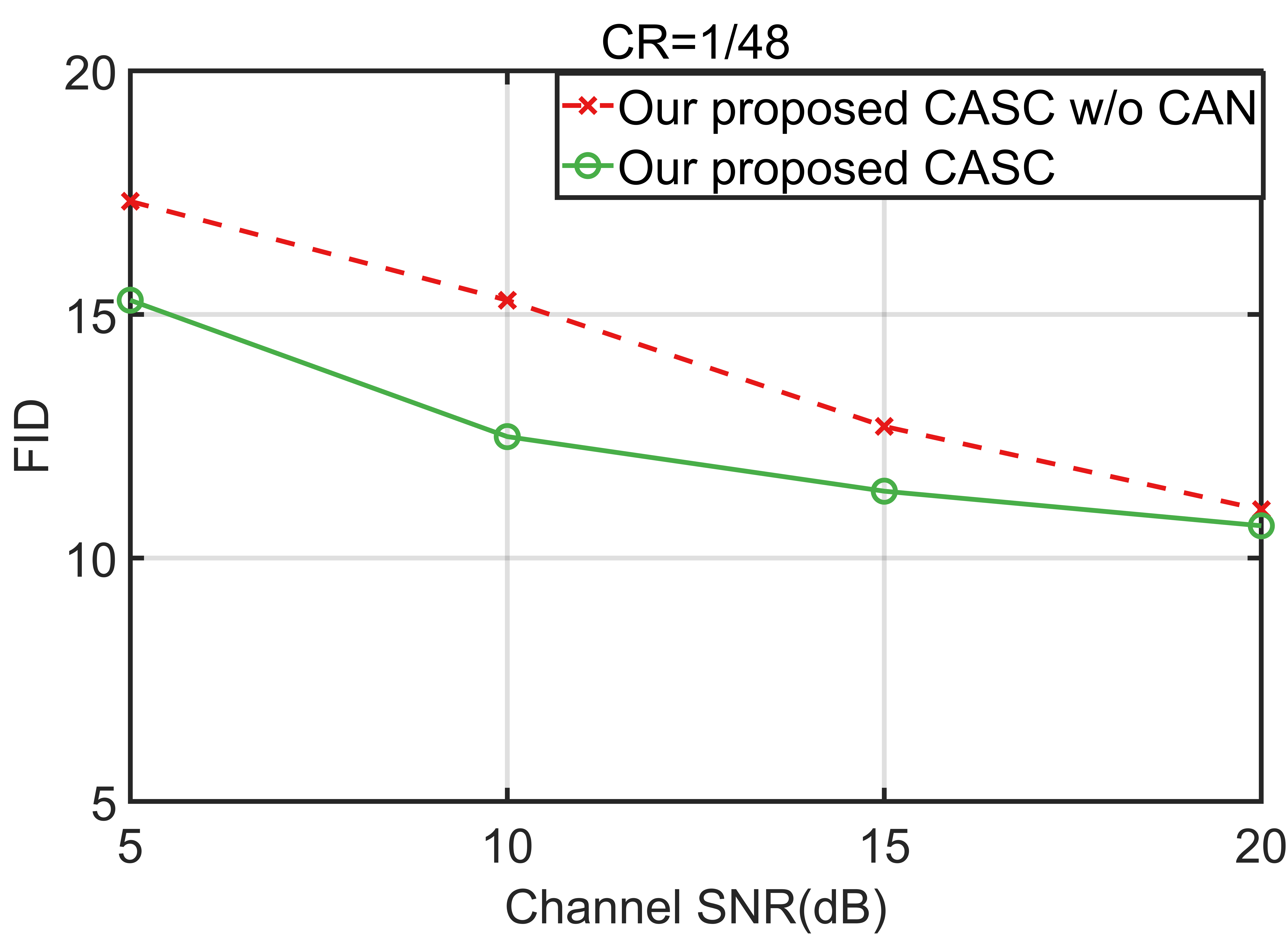}
\end{minipage}%
}%
\centering
\caption{Ablation experiment results at different channel SNRs with a CR of 1/48.}
\label{fig.ablation}
\vskip -0.1in
\end{figure*}


\subsection{Implementation Details}

Our experiments utilize the CIFAR-10 dataset \cite{krizhevsky2009learning},
where all images are $32 \times 32$ RGB images. 
For the autoencoder, $CH$ is set to 128, and $M$ is set to 2.
The length of the \textit{real-valued} condition signal $\textbf{C}$ is $d$, which determines the compression ratio (CR). 
The compression ratio is defined as $CR=d/2n$. 
In this paper, we train CASC with two different CRs, 1/48 and 1/96. 
The training of CASC is performed on the AWGN channel, and the channel signal-to-noise ratio (SNR) falls within the set $\rm SNR\in\{5,10,15,20\}dB$. 
The inference time steps $T$ is set to 1000.
We compare the performance of our proposed CASC with deep JSCC optimized using the mean square error (MSE) loss function, deep JSCC optimized using the LPIPS \cite{zhang2018unreasonable} loss function, and DiffSC method proposed in \cite{jiang2024diffsc}. 
The evaluation metrics used are peak signal-to-noise ratio (PSNR), LPIPS, and FID \cite{heusel2017gans}, where PSNR is a distortion metric, and LPIPS and FID are perceptual metrics.

\subsection{Performance Comparison}



Figs.~\ref{fig.CR=1/48} and ~\ref{fig.CR=1/96} show the performance comparison when the CR is 1/48 and 1/96, respectively. 
At low SNRs, the PSNR and LPIPS performance of CASC is comparable to that of the two Deep JSCC methods. 
However, as the SNR increases, CASC outperforms the two Deep JSCC methods in both PSNR and LPIPS performance.
The higher the SNR, the greater the advantage of CASC, which is attributed to the reduced noise interference in the condition signal, allowing for more accurate generation.
%
Furthermore, the FID performance of CASC significantly outperforms that of the two Deep JSCC methods at all SNRs.
The performance gap between CASC and the two Deep JSCC methods is even more pronounced when the CR is lower.
%
These results demonstrate the advantages of CASC in scenarios with limited bandwidth and favorable channel conditions, particularly in terms of perceptual performance.
In addition, CASC is comparable to DiffSC method \cite{jiang2024diffsc} in terms of perceptual performance, though it is slightly inferior to DiffSC method in PSNR performance. 
Notably, CASC offers the advantage of being lightweight, which will be discussed in the next subsection. 

\subsection{Inference Speed Comparison}


To demonstrate the inference speed advantage of CASC, we test the inference times of both CASC and DiffSC on a single NVIDIA RTX A6000 GPU under identical settings. 
The batch size is set to 256, and the single-image inference time of each system is subsequently calculated. 
The results are presented in Table~\ref{table1}.

\begin{table}[htbp]
\caption{The single-image inference time of CASC and DiffSC \cite{jiang2024diffsc} on a single NVIDIA RTX A6000 GPU.}
\label{table1}
\centering
\begin{tabularx}{0.4\textwidth}{>{\centering\arraybackslash}X|>{\centering\arraybackslash}X} 
\hline
\multicolumn{2}{c}{Inference time per image (ms)} \\ \hline
CASC & DiffSC \cite{jiang2024diffsc} \\ \hline
145.82 \textbf{($\downarrow$ 51.7\%)} & 301.80 \\ \hline
\end{tabularx}
\vskip -0.1in
\end{table}


The single-image inference time of CASC shows a 51.7\% reduction compared to DiffSC, indicating that CASC is particularly advantageous in applications where efficiency is crucial.


%

\subsection{Visualization Comparison}


Fig.~\ref{fig.visual} shows the visual comparison when the SNR is 20dB and the CR is 1/48.
Due to the very low CR, the image reconstructed by Deep JSCC (MSE) is significantly blurred, while the image reconstructed by Deep JSCC (LPIPS) contains a large number of noisy pixels. 
The reconstructed images of the two Deep JSCC methods lose substantial semantic information and have poor visual quality.
In contrast, the image reconstructed by CASC is visually pleasant and almost identical to the source image, indicating that CASC significantly outperforms the two Deep JSCC methods in terms of the visual quality of the reconstructed image.

\subsection{Ablation Studies}

Fig.~\ref{fig.ablation} shows the ablation experiment results when the CR is 1/48, demonstrating the effectiveness of CAN. 
The results indicate that both the distortion performance and perceptual performance of CASC are superior to those of CASC without CAN, proving that CAN effectively controls the generation process and enhances the overall performance of CASC. 
Additionally, CAN is particularly effective at lower SNRs.

\section{Conclusion}


In this paper, we proposed CASC, a condition-aware semantic communication framework with latent diffusion models for reliable image transmission with high perceptual quality over AWGN channels. To reconstruct the source image with low resource overhead and high perceptual quality, the latent diffusion model (LDM) was incorporated into the decoding stage, using noisy latent codes as the condition signal. Additionally, we introduced a condition-aware neural network (CAN) that provided dynamic weights to the hidden layers of the LDM, further enhancing perceptual quality. 
%
Our proposed framework demonstrated promising results for the application of LDM in SemCom.


\bibliographystyle{IEEEtran}

\bibliography{myref}


\vspace{12pt}
\end{CJK}
\end{document}